\documentclass[aps,prx,twocolumn]{revtex4-1}
\usepackage{amsmath,amssymb,graphicx,color,bm,epstopdf}
\usepackage{braket}
\usepackage{enumitem}

\newlist{UR}{enumerate}{1}
\setlist[UR]{label=[M\arabic*]}

\begin{document}

\title{Quantum inverse iteration algorithm for programmable quantum simulators}

\author{Oleksandr Kyriienko}
\affiliation{Department of Physics and Astronomy, University of Exeter, Stocker Road, Exeter EX4 4QL, UK}
\affiliation{ITMO University, Kronverkskiy prospekt 49, Saint Petersburg 197101, Russia}
\affiliation{NORDITA, KTH Royal Institute of Technology and Stockholm University, Roslagstullsbacken 23, SE-106 91 Stockholm, Sweden}

\begin{abstract}
We propose a quantum inverse iteration algorithm which can be used to estimate the ground state properties of a programmable quantum device. The method relies on the inverse power iteration technique, where the sequential application of the Hamiltonian inverse to an initial state prepares an approximate groundstate. To apply the inverse Hamiltonian operation, we write it as a sum of unitary evolution operators using the Fourier approximation approach. This allows to reformulate the protocol as separate measurements for the overlap of initial and propagated wavefunction. The algorithm thus crucially depends on the ability to run Hamiltonian dynamics with an available quantum device. We benchmark the performance using paradigmatic examples of quantum chemistry, corresponding to molecular hydrogen and beryllium hydride. Finally, we show its use for studying the ground state properties of relevant material science models which can be simulated with existing devices, considering an example of the Bose-Hubbard atomic simulator.
\end{abstract}

\maketitle

\section{Introduction} 

Quantum computing offers the drastic speed up for certain computational problems, and has evolved as a unique direction in the theoretical information science~\cite{NielsenChuang}. However, the field of experimental quantum computing is yet at its infancy. The typical size of quantum chips for the reliable gate based quantum computation ranges from one to several tens of physical qubits, with the main limits posed by decoherence. Despite the imperfections, the algorithms of ever-increasing complexity were implemented on different platforms, with circuit depth exceeding a thousand gates~\cite{Barends2016,Martinez2016}, potentially allowing for quantum supremacy demonstration.

At the same time, the \emph{vox populi} of quantum engineers says that while experimental setups are developed and mastered rapidly, the theorists in the field lag behind. Whereas by now textbook examples of quantum algorithms with exponential and quadratic speed up for factoring and search serve as a great motivation \cite{NielsenChuang}, the estimates of gate counts are daunting, making them distant goals for the future fault tolerant quantum computers~\cite{Wecker2014}. The recent developments in this fast evolving field call for new short depth algorithms which can solve \emph{useful} problems in the era of noisy intermediate scale quantum (NISQ) devices~\cite{Preskill2018}.

One of the most promising directions for quantum computation is the field of quantum chemistry and materials~\cite{Wecker2014,Wecker2015}. Targeting the access to the ground state properties of molecules and strongly correlated matter, it can offer huge gain for various technological applications, for instance helping to find a catalyst for the nitrogen fixation~\cite{Reiher2016}. To date, different quantum theoretical protocols were developed, and several proof-of-principle experiments on various platforms were performed in the simplest cases. Examples include simulation of molecular hydrogen with linear optical setup~\cite{Lanyon2010}, superconducting circuits~\cite{OMalley2015,Kandala2017,Colless2018,Ganzhorn2018}, and trapped ions~\cite{Hempel2018}. Finally, the variational simulation for larger molecules (LiH and BeH$_2$) were reported recently \cite{Kandala2017}. From the material science perspective, the use of cold atom quantum simulators has shown great promise, where simulations of Fermi-Hubbard lattice dynamics~\cite{Mazurenko2017}, large scale quantum Rydberg chain~\cite{Bernien2017} and Ising model~\cite{Zhang2017}, and two-dimensional many-body localization \cite{Choi2016} have been performed. However, in the latter cases the analog approach to simulation is taken, given an access to unitary dynamics, while precluding the study of ground state properties.

To access the ground state properties of a quantum chemical Hamiltonian, several routines can be used (see Refs.~\cite{OxfordRev,ZapataRev} for a review). First option corresponds to the quantum phase estimation algorithm (PEA)~\cite{Kitaev1997}, which exploits the unitary dynamics of the system controlled by register qubits. Although this algorithm is efficient, giving logarithmic error and polynomial gate scaling, its implementation requires substantial circuit depth for currently available circuits~\cite{OMalley2015}. Moreover, the controlled type of operations require the digitization of the circuit, thus complicating the use of analog quantum circuits for PEA.
Another approach is the adiabatic quantum computing, which was already applied to quantum chemistry problems~\cite{Babbush2014}. However, the required adiabaticity of dynamics typically results in the effectively long circuit depth.
Finally, an alternative route to quantum chemistry and materials is offered by hybrid-classical variational approaches which were proposed recently~\cite{Peruzzo2014}. They rely on term-by-term energy measurement for the prepared trial quantum state (ansatz) with consequent classical optimization, and are referred to as Variational Quantum Eigensolvers (VQE) \cite{McClean2016,OxfordRev,ZapataRev}. It can use a chemically inspired ansatz~\cite{McClean2016}, Hamiltonian variational ansatz~\cite{Wecker2015_VQE}, or rely on the variational imaginary time evolution~\cite{McArdle2018}. In this case the depth of the quantum circuit is greatly reduced, though at the expense of increased number of measurements, being favourable strategy for NISQ devices. For VQE the number of variational parameters scales as $O[(3N)^k]$, where $N$ is a number of qubits and $k$ represents an approximation order~\cite{McClean2016}. While for quantum chemistry applications $k=2$ suffices to give useful results, these approaches are yet to be tested for larger system sizes, where multi-variable optimization may raise the problems for genuine ground state estimation~\cite{McClean2018}.

In the following, we propose the quantum inverse iteration algorithm for the estimation of the ground state energy (GSE) of a quantum system. It is inspired by the classical inverse power iteration algorithm for finding the dominant eigenstate of the matrix, where the computationally demanding part of matrix inversion and multiplication is performed with a quantum circuit. Previously, a direct iteration approach was considered as a general purpose quantum algorithm~\cite{GeCirac2018}, aiming for a large scale fault tolerant implementation. Here, we present the protocol of the hybrid quantum-classical nature. It relies on performing quantum evolution for different propagation times and classical post-processing of the measured observables. The approach is applied to quantum chemistry examples (H$_2$ and BeH$_2$ molecules), showing favourable scaling with system parameters. Finally, when applied to the Bose-Hubbard quantum simulator, it allows to study its ground state properties, thus showing promise as a protocol for near-term quantum devices.
%
\begin{figure*}
\includegraphics[width=1.\textwidth]{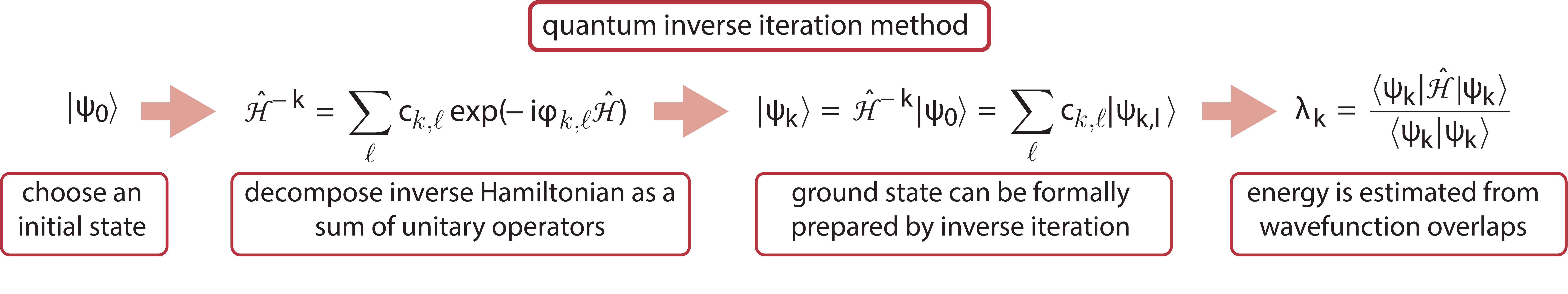}
\caption{Flowchart of the quantum inverse iteration algorithm. First, the initial product state is prepared and the inverse Hamiltonian operation is represented as a sum unitary evolution operators. Next, the iterated wavefunction can be formally obtained by applying the inverse. Finally, physical quantities (e.g. energy) are estimated as an expectation values of corresponding operator, and recast as a sum of wavefunction overlaps.}
\label{fig:sketch}
\end{figure*}       
%

\section{Protocol} 

We start by considering a generic interacting system, which can be described by the second quantized Hamiltonian. It can be written as sum of two-body and four-body parts
\begin{align}
\label{eq:H}
\hat{\mathcal{H}} = \sum_{ij} v_{ij} \hat{a}_i^\dagger \hat{a}_j + \sum_{ijkl} V_{ijkl} \hat{a}_i^\dagger \hat{a}_j^\dagger \hat{a}_k \hat{a}_l,
\end{align}
where $\hat{a}_i^\dagger$ ($\hat{a}_i$) can correspond to the fermionic or bosonic creation (annihilation) operators, and cover broad range of models. In the fermionic case, the Hamiltonian \eqref{eq:H} can describe the full configuration interaction problems in quantum chemistry, with operator $\hat{a}_j$ corresponding to molecular orbital $j$. Using the existing mappings between fermionic and spin-1/2 systems, one can rewrite \eqref{eq:H} in the form of a local Hamiltonian $\hat{\mathcal{H}}$ for interacting qubits which involves strings of Pauli operators. The task is then to find the lowest eigenvalue of the large matrix $\hat{\mathcal{H}}$, corresponding to GSE.

\subsection{Inverse iteration} We propose a procedure which can be seen as a quantum version of the inverse power iteration algorithm for finding the dominant eigenvalue of the matrix, represented by the inverse of the Hamiltonian matrix $\hat{\mathcal{H}}^{-1}$, which is treated as a dimensionless matrix in this section. Given that $\hat{\mathcal{H}}$ is invertible, the order of eigenvalues is reversed and, with the appropriate shift of the diagonal to make eigenvalues positive, the power iteration allows to find GSE. Namely, starting with an initial state $|\psi_{0}\rangle$ which has non-zero overlap with the sought ground state $|\psi_{\mathrm{gs}}\rangle$, by repetitive application of the inverted matrix one can prepare (unnormalized) state, $|\widetilde{\psi}_{k}\rangle = (\hat{\mathcal{H}}^{-1})^k|\psi_{0}\rangle$, such that $|\langle \psi_{\mathrm{gs}}|\psi_{k}\rangle|^2 < \epsilon$ for sufficiently large number of iterations $k \geq K$ \cite{Panju2011}, where $|\psi_{k}\rangle = |\widetilde{\psi}_{k}\rangle/ \Vert |\widetilde{\psi}_{k}\rangle \Vert$ (Fig.~\ref{fig:sketch}A). We note that generally this method has favourable logarithmic complexity in the iteration depth, being $K = \log[\epsilon \sin^{-2}(\theta_0) / (\lambda_{1} - \lambda_{n})]/[2\log(\lambda_2/\lambda_1)]$, where $\lambda_{1}$, $\lambda_2$, and $\lambda_n$ correspond to dominant, sub-dominant, and smallest eigenvalue of $\hat{\mathcal{H}}^{-1}$. Here $\sin^{2} \theta_0$ parametrizes the overlap between $|\psi_0 \rangle$ and $|\psi_{\mathrm{gs}}\rangle$, marking that convergence of the procedure depends on the initial guess, and generally can be made nonzero taking $|\psi_0 \rangle$ as a random state. 
%
While classical power iteration methods generally have good convergence in the number of iterations, the main caveat comes from the complexity scaling with the system size. The requirement for $K$ matrix multiplications leads to $O[K 2^{2N}]$ operations (for dense matrices), yielding exponential scaling. Even worse situation is for the inverse matrix algorithm, where an overhead comes from the $\hat{\mathcal{H}}^{-1}$ calculation, requiring extra $O[2^N]$ operations.

\subsection{Fourier approximation} In the following we show that we can exploit the iterative procedure with logarithmic iteration depth in $\epsilon$, while providing exponential speed up for the inverse Hamiltonian multiplication process. The latter comes from the approximation theory \cite{Sachdeva2013}, observing that the inverse can be represented as an integral $x^{-1} = \int_0^{+\infty} \exp(-x y)dy$, which by applying the trapezoidal rule can be written as a sparse sum of exponents. For quantum systems a similar idea was proposed in Ref.~\cite{Childs2017}, where Fourier approximation of the Hamiltonian inverse was presented as a double integral of the unitary propagator. This was further used to design an efficient solver for the quantum linear equation system problem~\cite{Harrow2009,Childs2017}. Here we extend the Fourier approximation to the $k$-th power of the inverse ($k \geq 1$), which formally reads
\begin{align}
\label{eq:H-k_int}
\hat{\mathcal{H}}^{-k} = \frac{i \mathcal{N}_k}{\sqrt{2\pi}}\int\limits_{0}^{+\infty}dy \int\limits_{-\infty}^{+\infty}dz z y^{k-1} \exp(-z^2/2) \exp(-i y z \hat{\mathcal{H}}),
\end{align}
and $\mathcal{N}_k$ is a normalization factor. The integral can be then discretized as 
\begin{align}
\label{eq:H-k_sum}
\hat{\mathcal{H}}^{-k} \approx &\frac{i \mathcal{N}_k}{\sqrt{2\pi}}\sum\limits_{j_y = 0}^{M_y-1}\Delta_y (j_y \Delta_y)^{k-1} \sum\limits_{j_z = -M_z}^{M_z} \Delta_z (j_z \Delta_z)  \times \\ \notag
& \exp[- j_z^2 \Delta_z^2 /2] \exp[-i (j_y \Delta_y) (j_z \Delta_z) \hat{\mathcal{H}}],
\end{align}
where $\Delta_{y,z}$ correspond to the discretization steps for integration variables, and $M_{y,z}$ represent cutoffs for integration. Notably, once applied to the physical Hamiltonian inverse, the discretization variable $\Delta_z$ remains dimensionless, while $\Delta_y$ bares the units of \emph{inverse energy}, serving akin to discrete time variable. The success of approximation \eqref{eq:H-k_sum} depends on the condition number of the Hermitian matrix $\hat{\mathcal{H}}$, given by the ratio of its largest to smallest eigenvalue, $\kappa = \lambda^{\hat{\mathcal{H}}}_{n} / \lambda^{\hat{\mathcal{H}}}_{1}$. Finally, Eq.~\eqref{eq:H-k_sum} can be conveniently redefined as 
\begin{align}
\label{eq:H-k_simp}
\hat{\mathcal{H}}^{-k} = \sum\limits_{\ell = 1}^{L_k} c_{k,\ell} \exp(-i \phi_{k,\ell} \hat{\mathcal{H}}) \equiv \hat{\mathcal{H}}^{-k}_{\mathrm{a}}, 
\end{align}
where we have rewritten the double summation in Eq. \eqref{eq:H-k_sum} using the superindex $\ell(j_y, j_z)$, $\phi_{k,\ell} = (j_y \Delta_y) (j_z \Delta_z)$ is a phase of evolution for parameters chosen to discretize $k$-th inverse, and $L_k = M_y (2M_z + 1)$. Here $c_{k,\ell}$ represent purely imaginary coefficients for the series, and $|c_{k,\ell}|$ define corresponding weights. The required size and number of discretization steps $\Delta_{y,z}$ and $M_{y,z}$ depends on $\epsilon$ and $\kappa$ (see Methods, section A, for the details). Importantly, they set the maximal evolution phase $\phi_{\mathrm{max}} = (M_y \Delta_y) (M_z \Delta_z)$, which serves as an equivalent of the total gate count for analog quantum simulation.

As we need to prepare the approximate ground state by applying generally non-unitary operator $\hat{\mathcal{H}}^{-K}$ to the initial state $|\psi_0\rangle$, we shall either introduce an ancillary register to perform it, or properly account for the normalization of the resulting wavefunction. The former option is an excellent strategy for the future fault-tolerant devices, and has beneficial scaling (Methods, sec. A), while the latter is more suitable for programmable quantum simulators. In the following we present the strategy which can be applied to estimate the ground state properties by sequential evaluation of terms in the series. Similarly to VQE approaches, this relies on performing large number of measurements, and thus adds an extra complexity as compared to the generic implementation of the inverse operator. At the same time, term-by-term readout offers better resilience to errors where even imperfect procedure can yield reasonable GSE estimate for quantum simulators.

\subsection{Sequential energy estimation} Our final goal is to estimate observables of the system, provided that an approximate ground state is prepared. For any operator $\hat{\mathcal{A}}$ it can be retrieved from the measurement $A = \langle \psi_{\mathrm{gs}}|\hat{\mathcal{A}}|\psi_{\mathrm{gs}} \rangle/\langle \psi_{\mathrm{gs}}|\psi_\mathrm{gs} \rangle$, where the normalization is accounted for explicitly. In particular, we are interested in the calculation of the ground state energy $\lambda_{\mathrm{gs}} \approx \lambda_k$, choosing the operator $\hat{\mathcal{A}}$ as $\hat{\mathcal{H}}$. This amounts to measurement of Hamiltonian expectation value for $|\psi_k\rangle = \hat{\mathcal{H}}^{-k} |\psi_0\rangle$ in the form
\begin{equation}
\label{eq:lambda}
\lambda_k = \frac{\langle \psi_{k}|\hat{\mathcal{H}}|\psi_{k} \rangle}{\langle \psi_{k}|\psi_{k} \rangle}.
\end{equation}
We proceed by considering each propagated wavefunction separately, such that $\lambda_k$ can be related to wavefunction overlaps (see flowchart in Fig.~\ref{fig:sketch}). This is motivated by the Hamiltonian averaging procedure \cite{Romero2017} used in VQE to reduce the circuit depth at the expense of larger number of sequential measurements. Using Fourier expansion of the inverse Hamiltonian \eqref{eq:H-k_simp}, the estimated energy reads
\begin{align}
\label{eq:lambda_sum}
&\lambda_k^{\mathrm{(a)}} = \frac{\sum_{\ell,\ell'} \langle \psi_{k,\ell'}|\hat{\mathcal{H}}|\psi_{k,\ell} \rangle}{\sum_{\ell,\ell'} \langle \psi_{k,\ell'}|\psi_{k,\ell} \rangle} \\ \notag &= \frac{\sum\limits_{\ell, \ell '} c^*_{k,\ell '} c_{k,\ell} \langle \psi_0 | e^{-i(\phi_{k,\ell}-\phi_{k,\ell '}) \hat{\mathcal{H}}} \hat{\mathcal{H}} |\psi_0 \rangle}{\sum\limits_{\ell, \ell '} c^*_{k,\ell '} c_{k,\ell} \langle \psi_0 | e^{-i(\phi_{k,\ell}-\phi_{k,\ell '}) \hat{\mathcal{H}}} |\psi_0 \rangle}.
\end{align}
Note that expression \eqref{eq:lambda_sum} now includes overlaps between initial and evolved wavefunction for the fixed phase, which shall be calculated separately for the numerator (``energy'') and denominator (``norm'').
Finally, we note that the wavefunction overlap can be inferred using different approaches. One option corresponds to using the SWAP test \cite{Ekert2002,Higgott2018}, which represents a common digital strategy and requires system doubling. This can be conveniently realized in some near-term setups, being successfully demonstrated for cold atom lattices by many-body interferometry of two copies of a quantum state~\cite{Islam2015}. Additionally, in the Methods, section B, we describe an alternative approach which does not require extra qubits and relies on the measurements of observables, once the reference state for the system is chosen.

\section{Results: quantum chemistry applications}

In the previous sections we described the general algorithm and discussed its key properties, namely the scaling and sequential operation. To show its use for the ground state estimation and characterize the required resources for realistic problems, we apply it to quantum chemistry. There, new strategies are much required due to rapidly growing complexity with the number of qubits (orbitals) $N$.
\begin{figure*}
\includegraphics[width=1.\textwidth]{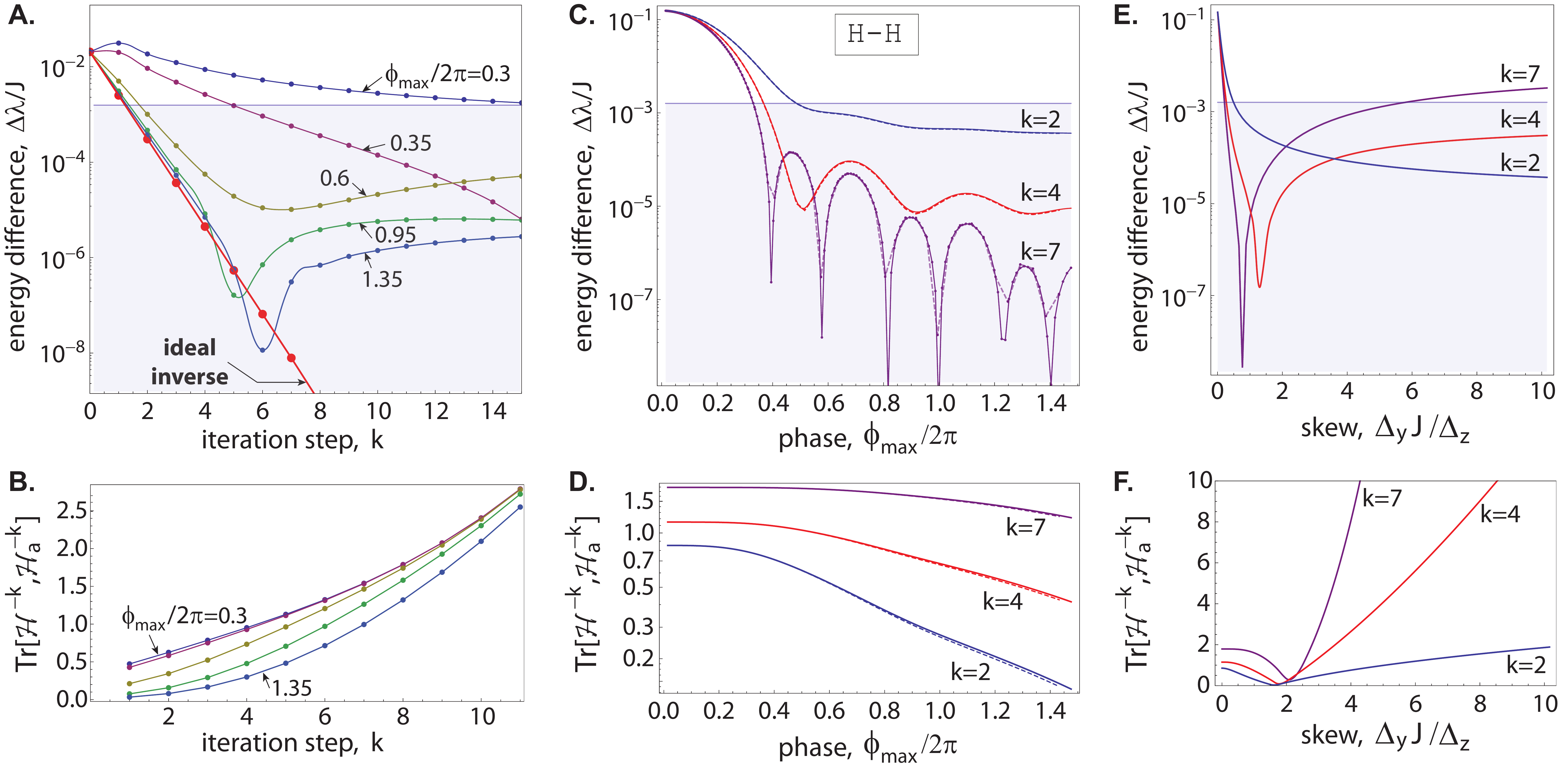}
\caption{Molecular hydrogen (H--H) example for benchmarking the quantum inverse power iteration ($N=4$ qubits). \textbf{A} Energy difference between the exact ground state and quantum inverse iteration estimate, shown as a function of the iteration step $k$ for different maximal phases of Fourier approximation (log scale). Solid red line shows the result for the ideal inverse iteration. Blue shaded area corresponds to the chemically precise estimate (same in \textbf{C,E}). \textbf{B} Trace distance $\mathrm{Tr}[\hat{\mathcal{H}}^{-k},\hat{\mathcal{H}}^{-k}_{\mathrm{a}}]$ between the ideal and approximate inverse operators shown as a function of iteration step number for different phases. \textbf{C} Energy difference $\Delta\lambda$ vs maximal propagation phase at different $k$ (log scale). \textbf{D} Trace distance $\mathrm{Tr}[\hat{\mathcal{H}}^{-k},\hat{\mathcal{H}}^{-k}_{\mathrm{a}}]$ vs phase for $k=2,4,7$. \textbf{E}, \textbf{F} Energy difference (\textbf{E}) and trace distance (\textbf{F}) plotted for the fixed maximal phase of $\phi_{\mathrm{max}}/2\pi = 0.92$, but different arrangement of the approximation grid defined by the skew parameter $\Delta_y J/\Delta_z$. Several iteration steps $k=2,4,7$ are depicted.}
\label{fig:appr}
\end{figure*}
%

\subsection{Molecular hydrogen}

We start with by now the standard example of molecular hydrogen, H$_2$. As a test task we consider the \textit{spinful} case. This allows to examine the protocol for a system of higher complexity ($N=4$), comparable to lithium hydrate four-qubit simulation considered in Ref. \cite{Kandala2017}. The details of mapping of quantum chemical structure into qubits are presented in the Methods, section~C. In the following we work with four-qubit molecular hydrogen Hamiltonian $\hat{\mathcal{H}}_{\mathrm{H}_2}$, with all eigenenergies shifted to positive values. Starting from the Hartree-Fock (HF) energy $\lambda_0$, the task is to estimate GSE $\lambda_{\mathrm{gs}}$, using the protocol described in the preceding section. This shall be done within the chemical precision $\epsilon$, which is equal to $\epsilon = 0.0016$~Hartree, and thus defines the relevant cut-off for the iteration procedure.

We start by benchmarking the inverse power procedure in its general form, and define how many iteration steps one needs to come close to the ground state. For this, we first perform the inverse Hamiltonian iteration in the ideal setting, assuming that an exact inverse is known. Then, we compare it to the quantum inverse iteration, which uses the Fourier approximation \eqref{eq:H-k_simp}. GSE is estimated using the measurement of propagated and initial wavefunction overlaps. To quantify the performance two characteristics are employed. The first, and the most natural one, corresponds to the difference between estimated energy value $\lambda_k$ and true GSE $\lambda_{\mathrm{gs}}$, being $\Delta \lambda/J \equiv (\lambda_k - \lambda_{\mathrm{gs}})/J$. It allows to observe the convergence and provides an indication of how well the procedure works for a given system. The second quantity corresponds to the trace distance between an idealized inverse iteration matrix $\hat{\mathcal{H}}^{-k}$ and its approximation $\hat{\mathcal{H}}^{-k}_a$, defined as a half of trace norm for the difference of two matrices. It reveals the actual success of mimicking the ideal inverse in full generality. At the same time, this is the quantity which cannot be straightforwardly observed in the experiment, and only serves for the analysis.
\begin{figure}[t!]
\includegraphics[width=1.\columnwidth]{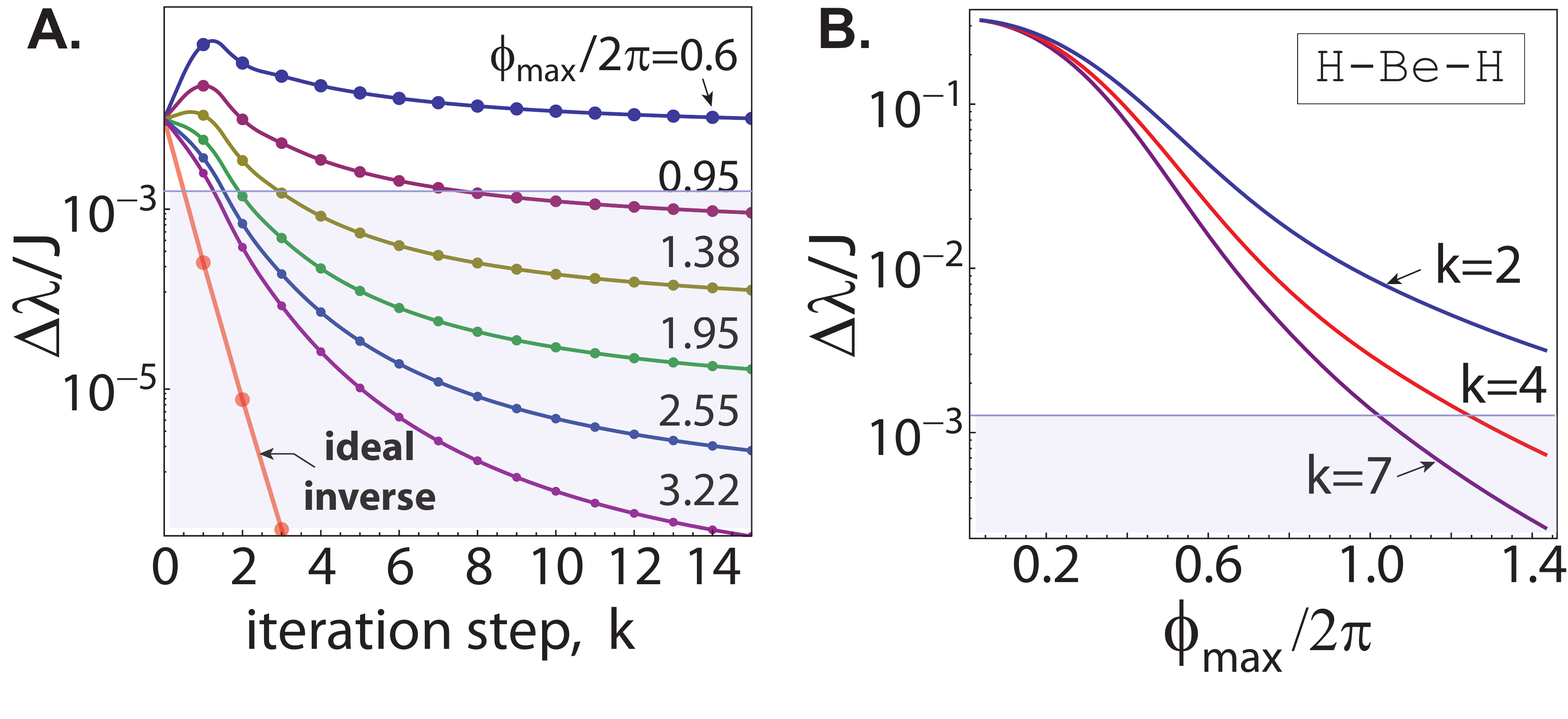}
\caption{Beryllium hydrate (H--Be--H) ground state energy estimation ($N=8$ qubits). \textbf{A} Energy difference for true GSE and quantum inverse iteration shown for different iteration steps $k$ and maximal propagation phases $\phi_{\mathrm{max}}$ (log scale). The red line corresponds to an ideal iteration procedure. \textbf{B} Same energy difference shown as a function of maximal phase at $k=2,4,7$. In both panels the blue shaded area corresponds to the chemically precise estimate.}
\label{fig:BeH2}
\end{figure}

The results of the inverse power iteration for molecular hydrogen Hamiltonian $\hat{\mathcal{H}}_{\mathrm{H}_2}$ are shown in Fig. \ref{fig:appr}A as a function of iteration step $k$. The ideal version of inverse iteration is plotted in red and reveals exponential convergence to GSE. The chemical precision is achieved already at the second step of the iteration, as depicted by the blue shaded area starting at $\Delta\lambda = 1.6\times 10^{-3} J$. The idealized case is then compared to the quantum inverse iteration procedure with combined measurement of wavefunction overlaps as stated in Eq.~\eqref{eq:lambda_sum}. Here, we assumed that the genuine unitary evolution with Hamiltonian $\hat{\mathcal{H}}_{\mathrm{H}_2}$ is run in the analog simulation fashion. The case of digital evolution with associated Trotterization technique and its benchmarking is considered in the Supplemental Material, where we also present the circuit scheme for digital evolution. The approximation was performed using equal number of steps $M_z = M_y = 30$, and the discretization values $\Delta_z = \Delta_y J$ were adjusted to match the maximal propagation phases of $\phi_{\mathrm{max}}/2\pi = J (M_y \Delta_y) (M_z \Delta_z)/2\pi = \{ 0.3, 0.35, 0.6, 0.95, 1.35\}$ (here, the propagation phase is taken to be dimensionless by absorbing energy unit prefactor $J$ from the Hamiltonian). The corresponding curves show the improvement of the quantum power iteration estimation for increasing number of iteration steps. The convergence rate also depends on the maximal phase of the propagation. For small phases (top curves in Fig.~\ref{fig:appr}A), the initial estimator does not give successful convergence, but comes closer to GSE for large $k$. As the propagation phase grows, the approximation $\lambda_k^{\mathrm{(a)}}$, starts to resemble the idealized iteration procedure. However, this only happens up to a certain value of $k$ past which the approximate energy grows, thus deviating from the ideal solution. From the point of view of process fidelity, the trace distance $\mathrm{Tr}[\hat{\mathcal{H}}^{-k},\hat{\mathcal{H}}^{-k}_{\mathrm{a}}]$ between ideal and approximate inverse operators increases monotonically with $k$ (Fig.~\ref{fig:appr}B). The increase of $\phi_{\mathrm{max}}$ allows to reduce $\mathrm{Tr}[\hat{\mathcal{H}}^{-k},\hat{\mathcal{H}}^{-k}_{\mathrm{a}}]$ at each $k$.
\begin{figure*}
\includegraphics[width=0.9\linewidth]{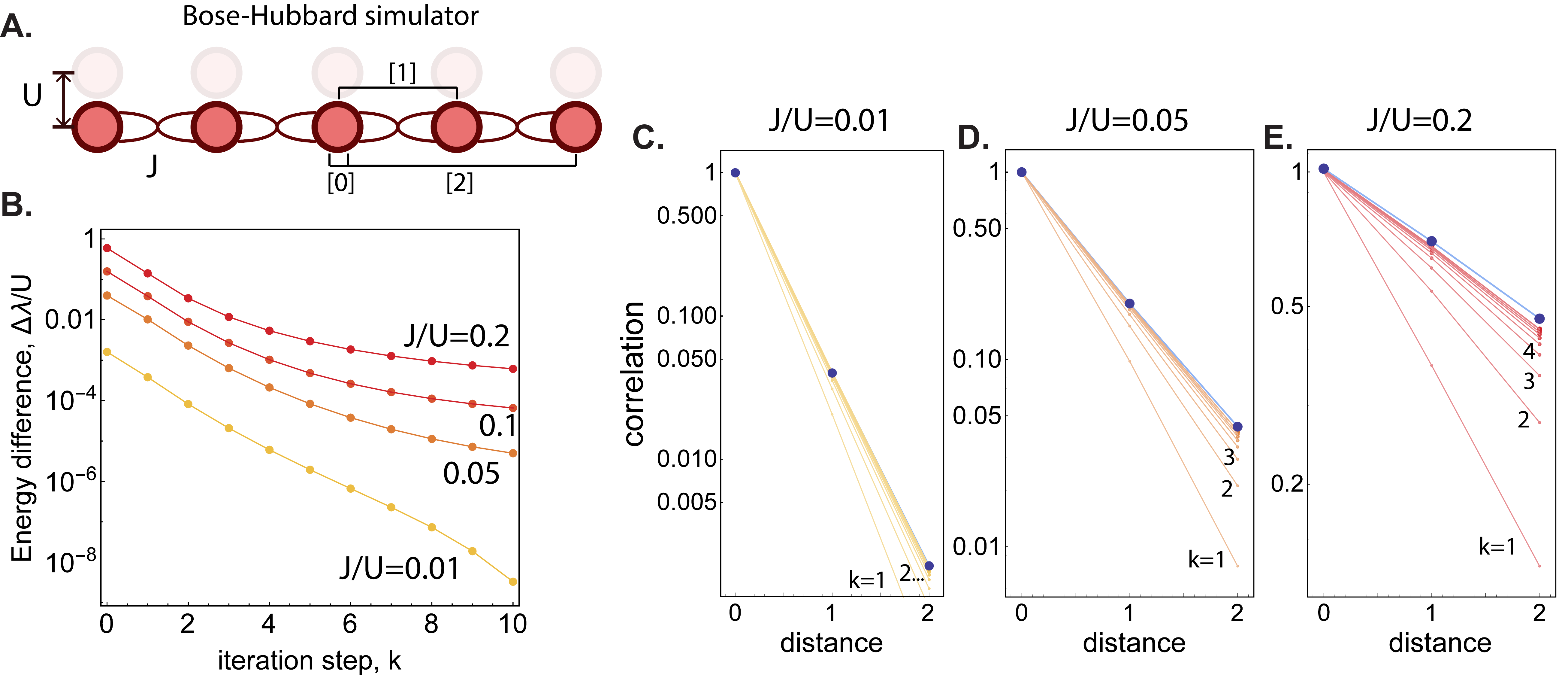}
\caption{Bose-Hubbard ground state energy estimation for a chain of five atoms. \textbf{A} Sketch of the system, which depicts coherent tunneling processes with the rate $J$ and the nonlinear interaction of strength $U$. Correlations are measured for a set of distances $r=[0,1,2]$. \textbf{B} Energy difference for the true GSE and quantum inverse iteration estimate are shown for different iteration steps $k$ (log scale). Several tunneling values are chosen, $J/U = 0.01,0.05,0.1,0.2$, capturing the transition between insulating and superfluid behaviour. \textbf{C,D,E} Correlation functions of the form $\langle \hat{a}_{c+r}^\dagger \hat{a}_{c} \rangle$ are calculated for the approximate ground state at increasing iteration step $k$, and for several values of $J/U$ (log scale). Correlations for true ground state are shown in blue (large dots and top curves).}
\label{fig:B-H}
\end{figure*}

The performance of the quantum inverse iteration procedure is further analyzed in Fig.~\ref{fig:appr}C,D where energy distance to ground state and trace distance are shown as a function $\phi_{\mathrm{max}}$ for several fixed iteration steps ($k=2,4,7$). The calculations were performed accounting for two different ways of arranging the phase. First, the approximation grid was fixed setting $M_y = M_z = 30$ while changing $\Delta_z = \Delta_y J$ (solid curves in Fig.~\ref{fig:appr}C,D). In the second case the fixed step size $\Delta_z = \Delta_y J = 0.05$ was combined with the increment of $M_{z,y}$ (dashed curves in Fig.~\ref{fig:appr}C,D). For both energy distance (Fig.~\ref{fig:appr}C) and trace distance (Fig.~\ref{fig:appr}D) we observe no difference between two approximation procedures, but clear indication of the importance of maximal propagation phase (time). For $\Delta\lambda$ one sees a non-monotonic dependence on $\phi_{\mathrm{max}}$ which starts with a decrease of the energy difference for increasing maximal phase ($\phi_{\mathrm{max}}/2\pi < 0.4$). At larger phases the dependence experiences pronounced dips (note the log scale), which are more visible for many iterations. Overall the difference remains well-within chemical precision and experiences saturation. When the trace distance is considered, one sees that success of the approximation monotonically improves with $\phi_{\mathrm{max}}$. At the same time, for fixed approximation parameters $\{M_{z,y},\Delta_{z,y}\}$ it is more difficult to represent inverse iteration operator faithfully, in-line with scaling analysis discussed in the Methods. Finally, the comparison of results in Fig.~\ref{fig:appr}C and D allows to suggest that non-monotonicity in the spectroscopic signatures can come from the particular structure of the Hamiltonian and the initial state, where certain phases might be preferable (i.e. not all elements of the Hamiltonian matrix contribute equally to the inverse iteration procedure).

To decide on the optimal way to approximate the inverse, we consider different discretization steps for $y$ and $z$ auxiliary variables, characterized by the skewness parameter $\Delta_y J/ \Delta_z$. The calculation is done for $M_z = M_y =30$ with the maximal phase fixed to $\phi_{\mathrm{max}}/2\pi = 0.92$. The results are shown in Fig.~\ref{fig:appr}E,F as a function of skew. The energy difference parameter shows that for approximating the inverse for small iteration numbers ($k=2$ curve in Fig.~\ref{fig:appr}E) larger skew factors are preferable, with $z$ variable requiring finer approximation. However, for increased iteration number the optimum flows to $\Delta_y J /\Delta_z \sim 1$ values, suggesting close-to-equal spacing can work well for varied $k$. Examining the trace distance, we see that in unbiased setting the skew ratio of $\Delta_y J/\Delta_z \sim 2$ is preferable.

Finally, the very important issue to address is an influence of noise on the operation of quantum inverse iteration protocol. For this we have performed the analysis including relevant dephasing processes which influence the estimate for overlaps (see details in the Supplemental Material). Although noise makes the estimation of energies at large iteration step $k$ less reliable, it is possible to estimate the energy within chemical precision using simple noise mitigation techniques.

\subsection{Beryllium hydride} 

To test the scalability of the approach, we consider a molecule of bigger size which requires larger Hilbert space simulation. For this, we choose to simulate beryllium hydride (BeH$_2$) in the full spinful version using $N=8$ qubits (see Methods, section D, for the details). 

We proceed in the same manner as for H$_2$ molecule, and quantify the operation of quantum inverse iteration procedure for BeH$_2$. The approximation parameters were chosen as $\Delta_y J = \Delta_z = 0.05$, with the number of discretization points $M_{y,z}$ adjusted accordingly to maintain maximal propagation phase. The results of the simulation are shown in Fig.~\ref{fig:BeH2}. The first plot (Fig.~\ref{fig:BeH2}A) shows that ideal iteration works well for the beryllium hydrate, with chemically precise GSE obtained already at $k=1$ iteration step. The Fourier approximation for the inverse at small phases does not reach required accuracy, while for increased iteration step number and $\phi_{\mathrm{max}}/2\pi > 1$ ground state estimate can be attained. Fig.~\ref{fig:BeH2}B shows this behavior as a function of phase for several representative $k$'s, and yields the same conclusion. The increase of the required propagation phase is attributed to the increased condition number for BeH$_2$ Hamiltonian matrix, being $39.2$ as compared to $3.38$ for H$_2$.

\subsection{Bose-Hubbard simulator} 

Finally, to provide an example where quantum inverse iteration algorithm can be largely beneficial, we consider the bosonic version of the Hubbard model (see sketch in Fig.~\ref{fig:B-H}A). The corresponding system Hamiltonian reads
\begin{align}
\label{eq:B-H_Hamiltonian}
\hat{\mathcal{H}}_{\mathrm{B-H}} = &-J \sum_{\langle i,j \rangle} (\hat{a}_i^\dagger \hat{a}_j + h.c.) + \frac{U}{2} \sum_i \hat{n}_i (\hat{n}_i - 1) \\ \notag &- \mu \sum_i \hat{n}_i,
\end{align}
where $\hat{a}_i^\dagger$ ($\hat{a}_i$) corresponds to the bosonic creation (annihilation) operator at lattice site $i$. Here, $\hat{n}_i = \hat{a}_i^\dagger \hat{a}_i$ corresponds to the number operator. The first term in the Hamiltonian \eqref{eq:B-H_Hamiltonian} describes the tunneling of bosons with rate $J$, which leads to their delocalization at the lattice. For simplicity we consider a one-dimensional lattice, and that tunneling happens only between neighbouring sites $\langle i,j \rangle$. The second term in \eqref{eq:B-H_Hamiltonian} denotes the contact interaction between bosons with strength $U$. The last term corresponds to the chemical potential $\mu$. Importantly, the Bose-Hubbard model corresponds to the paradigmatic example of a hard material science problem \cite{Childs}, and received considerable attention from both theoretical \cite{Freericks1994,White} and experimental perspective \cite{Bloch2012}. In particular, the model was shown to be easily solvable in the so-called Mott insulating regime where $U \gg J$ and ground state corresponds to the product state of one atom per site, $|\psi_{\mathrm{Mott}} \rangle = \prod_i |1\rangle_i $. However, going into the superfluid regime, where $J$ becomes comparable to $U$, the ground state is entangled, and it is generally difficult to study the low energy properties of this system.

We show that quantum inverse iteration algorithm can be applied to study the ground state properties of the Bose-Hubbard Hamiltonian in the wide range of parameters. For this, we consider an initial state corresponding to the Mott state,  $|\psi_{0} \rangle = |\psi_{\mathrm{Mott}} \rangle$, and perform the iteration as detailed in the ``Protocol'' section. The numerical results for the energy deviation are shown in Fig.~\ref{fig:B-H}B, where $\Delta\lambda/U \equiv |\lambda_k - \lambda_{\mathrm{gs}}|/U$ is plotted as a function of iteration step $k$. We consider $N=5$ lattice with $\mu /U = 0.5$, different values of tunneling rate $J$, and similarly to H$_2$ and BeH$_2$ examples the energies were adjusted by trivial overall shift $E_0$ to ensure positive eigenvalues. The Fourier approximation is performed using $M_y = M_z = 40$ and $\Delta_z = \Delta_y U = 0.075$. While for relatively small tunneling $J = 0.01 U$ the product state is a good ground state approximation, thus giving small energy deviation (lowest curve in Fig.~\ref{fig:B-H}B), for larger $J \sim 0.1 U$ the system enters a superfluid phase (critical value for $\mu = 0.5U$ approximately corresponds to $J_{\mathrm{crit}}/U \approx 0.13$ \cite{Freericks1994}). Despite the qualitatively different initial state, the algorithm allows to distill correct energy properties even for large $J$.

Going beyond the ground state energy estimation, quantum inverse iteration can be also used to measure the correlations in the ground state of the model. We considered the correlation function of the form $\langle \psi_k| \hat{a}_{c+r}^\dagger \hat{a}_{c} | \psi_k \rangle$, where $| \psi_k \rangle$ corresponds to the propagated state, and overall procedure is similar to one described in Eq.~\eqref{eq:lambda_sum}. We fix $c=3$ to be the central site in the lattice, and look for intra and intersite correlations with $r=0,1,2$. The results are shown in Fig.~\ref{fig:B-H}C,D,E, where in the Mott insulator regime correlation fall-off rapidly (C), while in the superfluid case the quantum inverse iteration method allows to restore genuine correlations (blue dots and curve) at increased $k$.

Importantly, the described Bose-Hubbard simulation can be performed experimentally using the available controllable devices \cite{Bloch2012,Islam2015}. First, it allows to run unitary dynamics for sufficiently long times and large particle numbers. Second, the many-body interference technique enables the convenient measurement of the wavefunction overlap, where part of the system is evolved and later is interfered with the initial copy \cite{Islam2015}. We note that our approach also bares similarities with recently introduced Quantum Virtual Cooling~\cite{Cotler2018}. This study demonstrated for reducing the temperature in half, while the inverse iteration offers the access to close-to-zero temperature. Finally, the intriguing possibility is an application of quantum inverse iteration to 2D models \cite{Choi2016} and Fermi-Hubbard model \cite{Mazurenko2017}, which shall be possible with improved interference techniques \cite{Cotler2018}.

\section{Discussion and conclusions}

We have presented the algorithm for the ground state energy (GSE) estimation of a quantum Hamiltonian. It is based on the iterative application of the Hamiltonian inverse to the initial state, and can be represented as a sum of unitary evolution operators. Targeting near-term quantum simulators, we described the protocol as a separate estimation of GSE contributions from the wavefunction overlap measurements. Then, the results from the quantum dynamical simulation are post-processed classically, and provide energy estimate for each iteration step. 

The algorithm was applied to several quantum chemistry examples, being molecular hydrogen and beryllium hydrate. Using the four-qubit H$_2$ simulator, we benchmarked the performance of iteration and inverse approximation, showing that the most valuable resource for GSE estimation is a maximally available time for unitary evolution. Both digital and noisy operation was considered, and found to be sufficient for a GSE calculation with chemical accuracy. 

As an outlook, we highlight that the approach can be beneficial for analog quantum simulators such as cold atoms lattices~\cite{Choi2016}, Rydberg atom simulators~\cite{Bernien2017}, trapped ions~\cite{Zhang2017}, and superconducting devices~\cite{Schuster2018}. For instance, the analog-type fermionic quantum chemistry simulator~\cite{Cirac2018} would be much valued for the task. Future applications also include material science problems, with the main target being Fermi-Hubbard model~\cite{Mazurenko2017}. For instance, we note that recently proposed approach of Quantum Virtual Cooling~\cite{Cotler2018}, which was experimentally applied to Bose-Hubbard model, has similar iterative structure and requires interferometric measurements. This poses the question of connection between the measurement-based cooling scheme and the dynamic protocol described in the current study.

\setcounter{equation}{0}
\setcounter{figure}{0}
\renewcommand{\theequation}{M\arabic{equation}}
\renewcommand{\thefigure}{M\arabic{figure}}

\small

\section*{Methods}

\subsection{Scaling and fault-tolerant implementation}

In this section we consider the scaling for the quantum inverse iteration algorithm. For $k=1$ it was shown in Ref.~[M1] that the inverse Hamiltonian can be approximated up to an error $\epsilon$ setting the discretization steps to $\Delta_y = \Theta (\epsilon/\log(\kappa/\epsilon))$, $\Delta_z = \Theta (1/\kappa \log(\kappa/\epsilon))$, and summing up to $M_y = \Theta ( \log(\kappa/\epsilon)\kappa/\epsilon)$, $M_z = \Theta (\kappa\log(\kappa/\epsilon))$ ($\kappa$ is a coordination number). The maximal evolution phase then scales as  $\phi_{\mathrm{max}} = (M_y \Delta_y) (M_z \Delta_z) = O( \kappa \log(\kappa/\epsilon))$, and is an equivalent of the total gate count for analog quantum simulation. Generalizing the result to $k$-th inverse iteration, the upper limit on the maximal required phase shall be multiplied by $K$, where truncation of inverse iteration introduces an additional error. In the study we considered particular examples, and quantified the validity of Fourier approximation as a function of $\{ M_y,\Delta_y,M_z,\Delta_z \}$ parameters.

Considering the fault-tolerant implementation, the approximate ground state can be prepared by applying generally non-unitary operator $\hat{\mathcal{H}}^{-K}$ to the initial state $|\psi_0\rangle$ using an ancillary qubit register. One possible option here is the amplitude amplification approach [M2] which addresses the task of implementation of the sum of unitary operators, of the same type as the one in Eq. \eqref{eq:H-k_simp}. Moreover, since we also require simulation of Hamiltonian dynamics for $\exp(-i \phi_{k,\ell}\hat{\mathcal{H}})$, which may be not accessible in analog-type simulation, the subsequent use of Hamiltonian simulation~[M3] or qubitization~[M4] methods would lead to favourable resource scaling. The algorithm will require $O(\log(L) \log(c\phi_{\mathrm{max}}/\epsilon)/\log\log(c\phi_{\mathrm{max}}/\epsilon))$ auxiliary qubits ($c \equiv \sum_\ell |c_\ell|$) and same order of controlled unitaries. 
This scaling can be compared to the iterative modification of the quantum phase estimation procedure (IPEA), based on a small fixed register~[M5] or a single auxiliary qubit~[M6]. The latest represents conceptually the closest algorithm to the one described in the paper, and thus will serve as benchmark. The complexity of IPEA was discussed in Ref.~[M6], showing the requirement of $O[\log(\epsilon) \log(\log(\epsilon)/\epsilon)]$ phase iterations to approach an error of $\epsilon = 2^{-m}$ (energy is rescaled such that $\Vert \hat{H} \Vert < 2 \pi$, and $m$ is the number of relevant bits of precision, typically limited to $<20$ for quantum chemistry applications). Each $k$-th IPEA step then requires implementation of the c-$U^k$ operation, defined as implementation of $(e^{-i \hat{H}})^k$, controlled on the register qubit. This leads to $O[N^4 \log(\epsilon) \log(\log(\epsilon)/\epsilon)]$ gate count, comparable to the inverse iteration procedure described above.

Given the favorable scaling, the cost of the general purpose quantum inverse iteration algorithm can be small for future large scale quantum devices. However, generally there is no simple procedure to perform c-$U$ operation, and it requires decomposition into a set of universal gates or multi-layer swap technique~[M7]. This enlarges the actual circuit depth (while being polynomial), and precludes the implementation of $U=\exp(-i\phi \hat{H})$ in analog fashion. Therefore, we target programmable devices with possible analog-type implementation and use sequential estimation strategy described in the main text.


\subsection{Overlap measurement}\label{sec:overlap} To measure the overlap between the evolved and initial wavefunction, we propose to exploit a single eigenstate $|\psi_{\mathrm{R}}\rangle$ of the system as a reference, and measure the overlap with respect to its energy $\lambda_{\mathrm{R}}$ (usually set to zero). This nicely fits the task of GSE estimation for the fermionic Hamiltonian, as its Hilbert space includes a vacuum state with no fermions present (unless space reduction procedure was performed). Similar technique was used for extracting spectroscopic signatures of photon localization~[M8], and the same approach was applied for the measurement of the density of states for the many-body system from the random state evolution~[M9]. The main steps for the measurement are as follows. The task is formulated as finding $\langle \psi_0 | \psi_0(t) \rangle$, where $|\psi_0\rangle$ is the initial state (typically corresponding to the Hartree-Fock solution). The state $|\psi_j(t) \rangle = \hat{\mathcal{U}}(t) |\psi_j \rangle$ is a time-propagated state with some unitary $\hat{\mathcal{U}}$ defined by the expansion. The HF state can be prepared from the reference $|\psi_{\mathrm{R}}\rangle$ (vacuum or other product state) using the product of local operators, and we note that these states are orthogonal. Then, the overlap probability is measured as an expectation value of the operator $\hat{M}_0 = |\psi_0\rangle \langle \psi_0|$ for time-evolved wavefunction, which reads 
\begin{align}
\label{eq:M_0}
\mathrm{Tr}\{ \hat{M}_0 |\psi_0(t)\rangle \langle \psi_0(t)| \} = |\langle \psi_0 | \psi_0(t) \rangle|^2 =: |\mathcal{O}_{0,t}|^2.
\end{align}
Next, the superposition of the vacuum and initial state shall be prepared as $|\psi_+ \rangle = (|\psi_{\mathrm{R}}\rangle + |\psi_0\rangle)/\sqrt{2}$ and evolved to $|\psi_+(t)\rangle$. Its overlap probability is measured as an expectation value of $\hat{M}_+ = |\psi_+\rangle \langle \psi_+|$ operator. This can be written as
\begin{align}
\label{eq:M_+}
&\mathrm{Tr}\{ \hat{M}_+ |\psi_+(t)\rangle \langle \psi_+(t)| \} = |\langle \psi_+ | \psi_+(t) \rangle|^2 \\ &= \frac{1}{4} \left( 1 + |\mathcal{O}_{0,t}|^2 + 2 \mathrm{Re}\{ \mathcal{O}_{0,t} e^{-i\lambda_{\mathrm{R}} t} \} \right)  =: |\mathcal{O}_{+,t}|^2,
\notag
\end{align}
and provides an information about real and imaginary parts of $\mathcal{O}_{0,t}$. The same procedure is performed for measuring an expectation value of the operator $\hat{M}_i = |\psi_i \rangle \langle \psi_i|$, where $|\psi_i \rangle = (|\psi_{\mathrm{R}}\rangle + i|\psi_0\rangle)/\sqrt{2}$. An additional information is gained with
\begin{align}
\label{eq:M_i}
&\mathrm{Tr}\{ \hat{M}_i |\psi_+(t)\rangle \langle \psi_+(t)| \} = |\langle \psi_i | \psi_+(t) \rangle|^2 \\ &= \frac{1}{4} \left( 1 + |\mathcal{O}_{0,t}|^2 - 2 \mathrm{Im}\{ \mathcal{O}_{0,t} e^{-i\lambda_{\mathrm{R}} t} \} \right)  =: |\mathcal{O}_{i,t}|^2,
\notag
\end{align}
and both real and imaginary part of $\mathcal{O}_{0,t}$ can be found for the known reference $\lambda_{\mathrm{R}}$ from the system of Eqs.~\eqref{eq:M_0}-\eqref{eq:M_i} as
\begin{align}
\label{eq:Re_O}
\mathrm{Re}\{ \mathcal{O}_{0,t} \} &= \Big[2|\mathcal{O}_{+,t}|^2 - (|\mathcal{O}_{0,t}|^2 + 1)/2 \Big] \cos(\lambda_{\mathrm{R}} t) \\ \notag &-\Big[2|\mathcal{O}_{i,t}|^2 - (|\mathcal{O}_{0,t}|^2 + 1)/2 \Big] \sin(\lambda_{\mathrm{R}} t),\\
\label{eq:Im_O}
\mathrm{Im}\{ \mathcal{O}_{0,t} \} = &-\Big[2|\mathcal{O}_{i,t}|^2 - (|\mathcal{O}_{0,t}|^2 + 1)/2 \Big] \cos(\lambda_{\mathrm{R}} t) \\ \notag & -\Big[2|\mathcal{O}_{+,t}|^2 - (|\mathcal{O}_{0,t}|^2 + 1)/2 \Big] \sin(\lambda_{\mathrm{R}} t).
\end{align}
Note that we are mostly interested in the real part of the sought overlap $\mathrm{Re}\{ \langle \psi_0 | \psi(t) \rangle \}$, as both the ``norm'' and ``energy'' terms of $\lambda_k$ are real, and imaginary parts of the overlap cancel out~[M10]. Thus it is also possible to deduce the real part indirectly as
\begin{align}
\label{eq:Re_O_alt}
|\mathrm{Re}\{ \mathcal{O}_{0,t} \}| = \sqrt{|\mathcal{O}_{0,t}|^2 - \mathrm{Im}\{ \mathcal{O}_{0,t} \}^2},
\end{align}
and its sign can be inferred from the measurement in Eq.~\eqref{eq:Re_O}. We note that in principle these are two related ways to estimate the real part of the overlap, corresponding to Eqs. \eqref{eq:Re_O} and \eqref{eq:Re_O_alt}. While being equivalent in the noiseless case, the effects of decoherence make two approaches distinct (see Supplemental Material for the details). For the practical purposes we thus refer to the overlap estimate in \eqref{eq:Re_O} as \emph{direct} estimation approach and refer to \eqref{eq:Re_O_alt} as the \emph{indirect} estimation.

Physically, the measurement procedure resembles the Bell-type measurement, where in the simple case of two-qubits a single CNOT operation and the Hadamard gate are required. For the larger system the measurement is generalized to GHZ-type, and consequently requires increasing number of two qubit operators, which depends on how much an initial HF state is different from the reference state.

Finally, we note that direct measurement is possible in atomic setups, where many-body interferometry is applied to two copies. This can be used in the analog circuits where system size doubling plays lesser role, and can compensate the absence of fully addressable individual gate operation.

\subsection{Molecular hydrogen Hamiltonian}

The fermionic Hamiltonian $\hat{\mathcal{H}}_{\mathrm{H}_2}$ is first written in the form~\eqref{eq:H}, where coefficients $v_{ij}$ and $V_{ijkl}$ are calculated by conventional quantum chemistry methods. Here, we exploited the OpenFermion package for Python [M11], which allows to extract the interfermionic interactions for four Gaussian orbitals fit via STO-3G method and perform the fermions-to-qubits transformation. For the small $N=4$ system we have chosen to use the Jordan-Wigner transformation, although other options may be used as the system size increases. Specifically, we consider the bond length for H$_2$ to be $0.7414$ (measured in \r{A}ngstr\"{o}m) and consider full excitation space. For concreteness, we provide the full form for the Hamiltonian, being
\begin{align}
\label{eq:H_H2_full}
&\hat{\mathcal{H}}_{\mathrm{H}_2} = \xi_0 + \xi_1 (Z_0 + Z_1) - \xi_2 (Z_2 + Z_3) + \xi_3 Z_0 Z_1 \\  \nonumber
& + \xi_4 (Z_0 Z_2 + Z_1 Z_3) + \xi_5 (Z_0 Z_3 + Z_1 Z_2) + \xi_6 Z_2 Z_3 \\  \nonumber &- \xi_7 (X_0 X_1 Y_2 Y_3 - X_0 Y_1 Y_2 X_3 - Y_0 X_1 X_2 Y_3 + Y_0 Y_1 X_2 X_3),
\end{align}
where $X_j, Y_j, Z_j$ denote Pauli matrices for qubit $j$. The coefficients read $\xi_0/J = -0.098864$, $\xi_1/J = 0.171198$, $\xi_2/J = 0.222786$, $\xi_3/J = 0.168622$, $\xi_4/J = 0.120545$, $\xi_5/J = 0.165867$, $\xi_6/J = 0.174348$, $\xi_7/J = 0.045322$. The energy scale $J$ for the actual H$_2$ Hamiltonian corresponds to Hartree units, while for the quantum simulator $J$ corresponds to the effective qubit coupling.  Throughput the text, we measure energy in units of $J$, and the time is measured in $J^{-1}$ units. The digital implementation for the unitary evolution with Hamiltonian~\eqref{eq:H_H2_full} is presented in the Supplemental Material. The Hartree-Fock (HF) solution for the problem is given by the approximate ground state $|\psi_0 \rangle= (\downarrow, \downarrow, \uparrow, \uparrow)^T$, and associated HF energy is $-1.116684J$. As required by the Fourier approximation approach, the reference energy is then shifted towards positive values by adding constant term equal to $E_0/J = 2.$, and we refer to the shifted Hamiltonian as $\hat{\mathcal{H}}_{\mathrm{H}_2}$ in the main text. Its HF energy is $\lambda_0 = 0.883316 J$ after the shift. The task is then to estimate the ground state energy $\lambda_{\mathrm{gs}}$ of the Hamiltonian $\hat{\mathcal{H}}_{\mathrm{H}_2}$, achieved by preparation of approximate ground state $|\psi_{k} \rangle$. This shall be done within the chemical precision $\epsilon$, which is equal to $\epsilon = 0.0016$~Hartree, and thus defines the relevant cut-off for the iteration procedure.

\subsection{Beryllium hydride Hamiltonian}
The molecular data structure of beryllium hydride (BeH$_2$) was generated using Psi4 quantum chemistry package~[M12] considering equal Be$-$H distances equal to 1.33 \r{A}ngstr\"{o}m. While generically described by six spin orbitals, we set lowest and second excited orbital to be occupied, and set multiplicity of unity, such that the ground state energy lies close to the full configurational space solution (STO-3G basis). The fermionic Hamiltonian is then obtained using OpenFermion package, and as in the case of H$_2$ the Jordan-Wigner transformation was used to rewrite it in the qubit form~[M12]. The problem then can be solved using $N=8$ qubits. The GSE from the exact diagonalization of original BeH$_2$ Hamiltonian reads $-1.806750$ Hartree, and analogously to the molecular hydrogen the Hamiltonian matrix is shifted by constant energy term of 2 Hartree. The product state corresponding to Hartree-Fock solution reads $|\psi_0 \rangle= (\downarrow, \downarrow, \uparrow, \uparrow, \uparrow, \uparrow, \uparrow, \uparrow)^T$, with associated energy for the shifted Hamiltonian being $\lambda_0/J = 0.203323$.

\normalsize

\begin{acknowledgments}
I would like to thank Anders S. S{\o}rensen for useful discussions and suggestions, and Bart Olsthoorn for technical advise and reading the manuscript.\\
\end{acknowledgments}

\section{Competing interests} The author declares that there are no competing interests.

\section{Author contributions} O.K. proposed the original idea, preformed the calculations, analyzed the results, and written the manuscript.

\section{Funding} The work was supported by the Government of the Russian Federation through the ITMO Fellowship and Professorship Program.

\section{Data availability} The authors declare that the data supporting the findings of this study are
available within the paper and its supplementary information file.


\begin{center}
\textbf{REFERENCES [Methods]}
\end{center}

\begin{UR}  

\item A. M. Childs, R. Kothari, and R. D. Somma, \textit{Quantum algorithm for systems of linear equations with exponentially improved dependence on precision}, SIAM Journal on Computing \textbf{46}, 1920 (2017).

\item G. Brassard, P. Hoyer, M. Mosca, and A. Tapp, \textit{Quantum Amplitude Amplification and Estimation}, Quantum Computation and Quantum Information, Samuel J. Lomonaco, Jr. (editor), AMS Contemporary Mathematics, 305:53-74 (2002); arXiv:0005055.

\item D. W. Berry, A. M. Childs, R. Cleve, R. Kothari, and R. D. Somma,\textit{ Simulating Hamiltonian Dynamics with a Truncated Taylor Series}, Phys. Rev. Lett. \textbf{114}, 090502 (2015).

\item Guang Hao Low and Isaac L. Chuang, \textit{Optimal Hamiltonian Simulation by Quantum Signal Processing}, Phys. Rev. Lett. \textbf{118}, 010501 (2017).

\item  A. Aspuru-Guzik, A. D. Dutoi, P. J. Love, M. Head-Gordon, \textit{Simulated Quantum Computation of Molecular Energies}, Science \textbf{309}, 1704 (2005).

\item M. Dobsicek, G. Johansson, V. Shumeiko, and G. Wendin, \textit{Arbitrary accuracy iterative quantum phase estimation algorithm using a single ancillary qubit: A two-qubit benchmark}, Phys. Rev. A \textbf{76}, 030306(R) (2007).

\item Xiao-Qi Zhou, T. C. Ralph, P. Kalasuwan, Mian Zhang, A. Peruzzo, B. P. Lanyon, and J. L. O'Brien, \textit{Adding control to arbitrary unknown quantum operations}, Nature Comm. \textbf{2}, 413 (2011).

\item P. Roushan, C. Neill, J. Tangpanitanon, V. M. Bastidas, A. Megrant, R. Barends, Y. Chen, Z. Chen, B. Chiaro, A. Dunsworth, A. Fowler, B. Foxen, M. Giustina, E. Jeffrey, J. Kelly, E. Lucero, J. Mutus, M. Neeley, C. Quintana, D. Sank, A. Vainsencher, J. Wenner, T. White, H. Neven, D. G. Angelakis, and J. Martinis,  \textit{Spectroscopic signatures of localization with interacting photons in superconducting qubits}, Science \textbf{358}, 1175 (2017).

\item O. Kyriienko and A. S. S{\o}rensen, \textit{Quantum protocol for the measurement of the density of states for quantum systems}, in progress.

\item This can be also shown formally if an exact form of expansion \eqref{eq:H-k_sum} is considered. The numerator of Eq. \eqref{eq:lambda_sum} can written term-by-term such that terms with equal $j_z = j_z '$ are considered, accompanied by associated exponents $e^{i\delta\phi \hat{\mathcal{H}}}$ and $e^{-i\delta\phi \hat{\mathcal{H}}}$. Similarly, the terms with $j_z = -j_z '$ can be grouped. Combined this leads to $\lambda_k \propto \langle \psi_0 | \cos(\delta\phi \hat{\mathcal{H}}) | \psi_0 \rangle$ dependence, while odd imaginary terms have vanished.

\item J. R. McClean \textit{et al.}, \textit{OpenFermion: The Electronic Structure Package for Quantum Computers}, arXiv:1710.07629 (2017).

\item R. M. Parrish \textit{et al.}, \textit{Psi4 1.1: An Open-Source Electronic Structure Program Emphasizing Automation, Advanced Libraries, and Interoperability}, J. Chem. Theory Comput. \textbf{13}, 3185 (2017).

\item A. Tranter, P. J. Love, F. Mintert, and P. V. Coveney, \textit{A comparison of the Bravyi-Kitaev and Jordan-Wigner transformations for the quantum simulation of quantum chemistry}, J. Chem. Theory Comput. \textbf{14}, 5617 (2018).

\end{UR}



\newpage


\renewcommand{\theequation}{S\arabic{equation}}
\renewcommand{\thefigure}{S\arabic{figure}}





\begin{center}

\textbf{Supplemental Material: Quantum inverse iteration algorithm for programmable quantum simulators}\\

~\\

by Oleksandr Kyriienko\\

\end{center}

In this Supplemental Material we provide the detailed analysis of the performance of quantum inverse iteration algorithm for the case of \emph{digitized} evolution, show the relevant digital operation scheme, and analyse the influence of noise.\\

\section{Trotterization}

As shown by the analysis in the main text (Sec. III.A), the success of the ground state estimation protocol depends on the implementation of unitary operators $\hat{U}(\phi)$ for various values of the phase. Thus it largely favors the analog-type simulation, where Hamiltonian evolution can be seen as a resource for ground state energy estimation. When this is not available, the corresponding unitary operator can be implemented approximately using the digital or Floquet strategies~\cite{Kyriienko2018_SM}. As the simplest approach for systems with limited resources we choose standard technique of Trotter expansion. In particular, we use the second-order expansion~\cite{Babbush2015_SM,Childs2018_SM}, where a single Trotter step unitary reads
\begin{align}
\label{eq:U_Trotter_step}
\hat{\mathcal{U}}_{\mathrm{Tr}}^{(1)}(\phi/N_{\mathrm{Tr}}) = \prod\limits_{m=1}^{M} e^{-i (\phi/2 N_{\mathrm{Tr}}) \hat{\mathcal{H}}_m } \prod\limits_{m=M}^{1} e^{ -i (\phi/2 N_{\mathrm{Tr}}) \hat{\mathcal{H}}_m },
\end{align}
and the full unitary reads
\begin{align}
\label{eq:U_Trotter}
\hat{\mathcal{U}}_{\mathrm{Tr}}(\phi) = \prod_{i=1}^{N_{\mathrm{Tr}}} \hat{\mathcal{U}}_{\mathrm{Tr}}^{(1)}(\phi/N_{\mathrm{Tr}}),
\end{align}
where $N_{\mathrm{Tr}}$ denotes the total number of Trotter steps. Here, the Hamiltonian is considered as a sum of terms $\hat{\mathcal{H}} = \sum_m^{M} \hat{\mathcal{H}}_m$ which can be implemented separately, and $M$ is the total number of terms. For the molecular hydrogen Hamiltonian it counts fourteen Pauli terms (see Methods, sec. C, of the main text), $M=14$. However, the first ten and the last four terms are mutually commuting, so effectively Trotterization procedure relies only on the application of two non-commuting parts $\hat{\mathcal{H}}_{\mathrm{Z's}}$ and $\hat{\mathcal{H}}_{\mathrm{strings}}$. We note that an exact realization of corresponding unitaries is platform-dependent, and thus shall be implemented specifically to the hardware capabilities. One potential implementation is offered in the end of this section.
\begin{figure}
\includegraphics[width=1.\columnwidth]{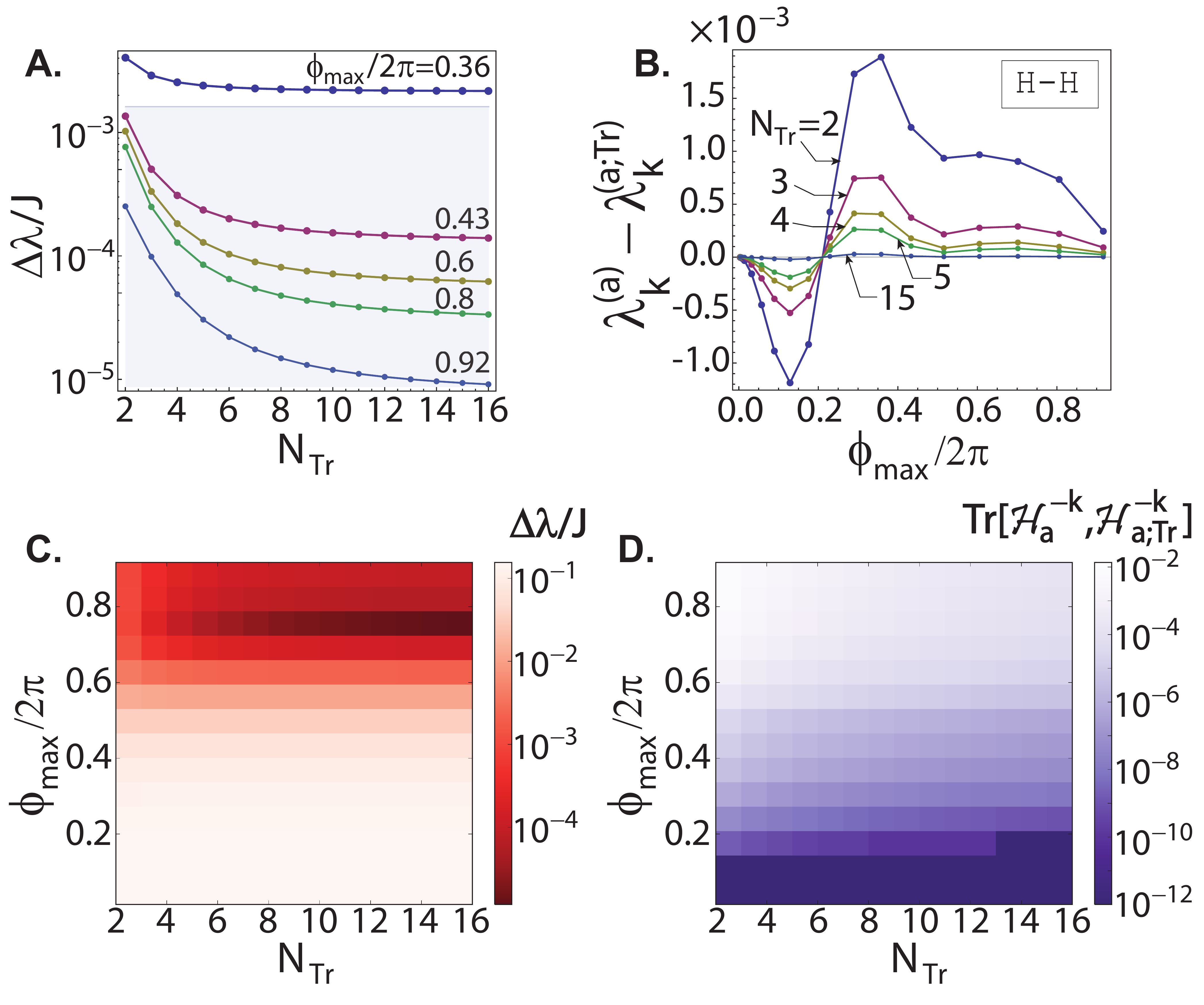}
\caption{Quantum inverse iteration procedure with Trotterized evolution for molecular hydrogen (H--H). \textbf{A} Difference between the true ground state energy and Trotterized quantum inverse iteration estimate shown for different numbers of Trotter steps $N_{\mathrm{Tr}}$ (log scale). Several values of maximal propagation phase are plotted. Iteration step is fixed to $k=4$ for each panel. Blue shaded area corresponds to the chemical precise estimate. \textbf{B} Energy difference between quantum inverse iteration estimates using continuous (analog) propagation and Trotterized implementation, plotted as a function of maximal propagation phase (linear scale with $\times 10^{-3}$ prefactor). Cases of different Trotter step number are shown ($N_{\mathrm{Tr}} = 2,3,4,5,15$). \textbf{C} Density plot for the difference between true ground state energy and Trotterized quantum inverse iteration estimate (absolute value, log scale) at different propagation phases and Trotter steps numbers. \textbf{D} Trace distance between Fourier approximations of inverse iteration operator ($k=4$) for an analog and digitized propagation, plotted at different propagation phases and Trotter steps numbers.}
\label{fig:Trotter}
\end{figure}
\begin{figure*}[t!]
\includegraphics[width=1.\textwidth]{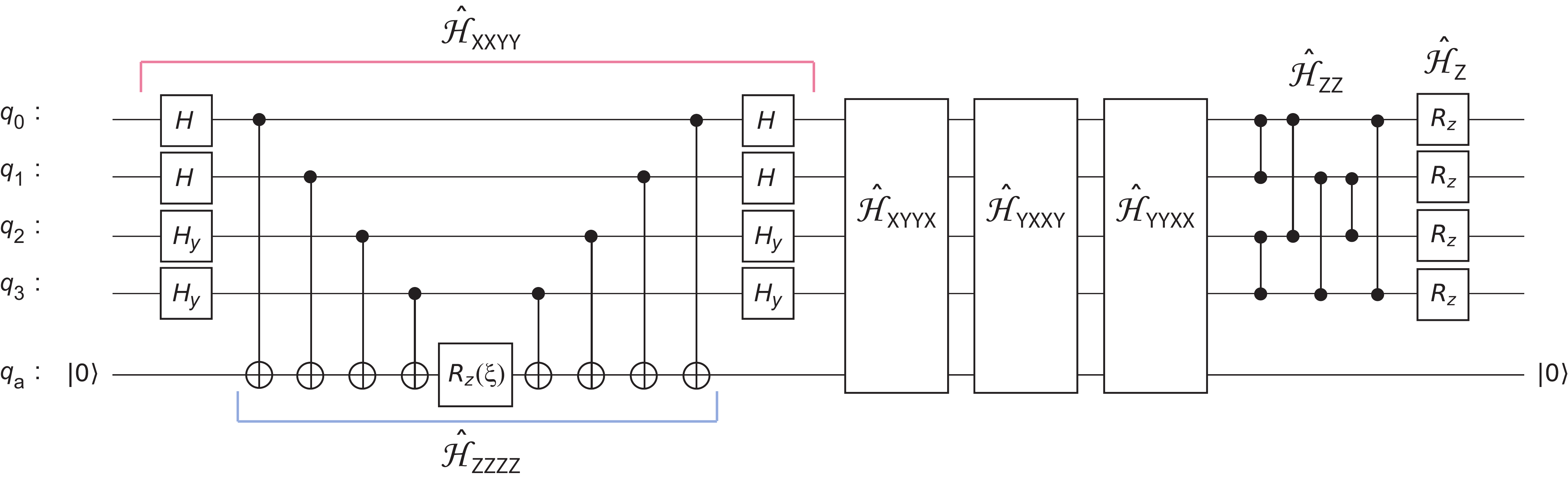}
\caption{The quantum circuit representation of the Trotter step required to simulate the dynamics of H$_2$ Hamiltonian in the digital form. Here, the set of CNOT operations allows to simulate the string-like Hamiltonian $\hat{\mathcal{H}}_{\mathrm{ZZZZ}}$, which is converted into required $\hat{\mathcal{H}}_{\mathrm{XXYY}}$-type strings with Hadamard gates. Finally, the Trotter step is completed by the layer of commuting CPHASE gates which simulate $\hat{\mathcal{H}}_{\mathrm{ZZ}} = \exp(-i\xi_{ij}Z_i Z_j)$ unitaries, and magnetic fields introduced by local Z rotations.}
\label{fig:circuit}
\end{figure*}

The performance of the quantum inverse iteration with approximate Trotterized unitary dynamics is shown in Fig.~\ref{fig:Trotter}. First, we plot the difference between estimated value and ideal ground state $\Delta \lambda$ as a function of Trotter step number, considering fixed iteration step $k = 4$ and several values of maximal propagation phase (Fig.~\ref{fig:Trotter}A). We see that for small $\phi_{\mathrm{max}}$ the estimate lies outside of chemical precision, as governed by Fourier approximation error for the inverse, while the dependence on Trotter step number is weak and shows quick convergence. For increasing phase (curves 0.43-0.92) the estimate improves, and even two Trotter steps may be sufficient, favoring NISQ device operation. At the same time, as expected for the Trotterization procedure, larger phase (evolution time) requires finer procedure, and shows stronger $N_{\mathrm{Tr}}$ dependence. The discrepancy between analog and digital unitary dynamics can be seen in Fig.~\ref{fig:Trotter}B, where the energy difference for both is plotted for various $N_{\mathrm{Tr}}$ as a function of phase. While generally the increase of Trotter step number leads to convergence between the two (see curve 15 staying close to zero at all phases), for small $N_{\mathrm{Tr}}$ the difference is a non-monotonous function of phase. In particular, at certain point the difference between $\lambda^{\mathrm{(a)}}_k$ and $\lambda^{\mathrm{(a;Tr)}}_k$ shrinks to zero, which can be related to structure of the Hamiltonian. Further characterization is performed showing spectral and process differences as a function of both $\phi_{\mathrm{max}}$ and $N_{\mathrm{Tr}}$. Similarly to analysis in Fig.~2 in the main text, the energy difference, shown in Fig.~\ref{fig:Trotter}C, reveals windows of largely reduced $\Delta\lambda$, which are more pronounced at large $N_{\mathrm{Tr}}$. Finally, the trace distance between inverse iteration operators with continuous and Trotterized dynamics demonstrate excellent convergence for small phases and large $N_{\mathrm{Tr}}$ (Fig.~\ref{fig:Trotter}D). No special phase points were spotted, as for the energy distance in Fig.~\ref{fig:Trotter}B.

Returning back to the question of physical implementation, we note that the major simulation difficulty corresponds to implementing the four-body interaction terms $\hat{\mathcal{H}}_{\mathrm{strings}}$. To do so, we can use classical technique (see Nielsen\&Chuang \cite{NielsenChuang_SM}, exercise 4.51) where a single ancillary qubit rotation $R_z(\xi)$ is used together with CNOT operations to simulate $\hat{\mathcal{H}}_{\mathrm{ZZZZ}}=\xi Z^{\otimes n}$ Hamiltonian (Fig.~\ref{fig:circuit}). Next, the Hadamard rotations $H$ and $H_y$ are used to change it into strings of $X$ and $Y$ operations, where $H = \exp(-i \pi (X+Z)/2\sqrt{2})$ and $H_y = \exp(-i \pi (Y+Z)/2\sqrt{2})$. The same procedure can be repeated to simulate other strings, and, importantly, no extra Trotterization error arises since the four strings commute. Alternatively, we note that same Hamiltonian structure is generated from the Floquet-type unitary where the set of $\pi$ X rotations is followed by the Ising Hamiltonian \cite{Kozin2019_SM}. Finally, the Trotter step is completed by introducing $\hat{\mathcal{H}}_{\mathrm{ZZ}}$ and $\hat{\mathcal{H}}_{\mathrm{Z}}$ terms from the H$_2$ Hamiltonian, which can be straightforwardly implemented with various platforms.

To estimate the feasibility of the physical realization, the fast convergence of the algorithm allows to reach chemical precision already with two Trotter steps. Given the increasing gate fidelities and improved connectivity of devices, where all-to-all two qubit gates can be performed on the square plaquette \cite{Google_SM}, the protocol requires $~150$ two-qubit and one-qubit gates. For $\epsilon = 10^{-3}$ noise level this can provide final state fidelities in $\sim 0.9$ range.

\section{Noisy operation, measurement, and error mitigation}

In the main text we considered the quantum processor to be noiseless. Namely, the deviation of estimated and true GSE was governed by the coherent errors associated to the insufficient iteration number or approximation grid. However, for near-term devices the important limitation corresponds to noise. To address this issue, we consider a realization of quantum iteration algorithm which accounts for the dephasing of qubits. The latter commonly serves as a dominant source of error for different quantum systems. Simultaneously, we provide the details for the measurement schedule, and show how the protocol can be separated into the quantum stage of measuring overlaps and classical post-processing stage.

The energy estimation procedure is performed as stated in Eq.~(6) of the main text. When Fourier approximation grid is taken to be the same for each iteration, the list of propagation phases $\phi_{\ell}$ does not depend on $k$. The summation over two indices $\ell, \ell '$ can be reduced to a single superindex $q(\ell, \ell')$, leading to the introduction of the phase difference list $\delta \phi_{q(\ell, \ell')} = \phi_{\ell }- \phi_{\ell'}$ and real-valued prefactor list $p_{k,q(\ell,\ell')} = c^*_{k,\ell '} c_{k,\ell}$. The evolved wavefunction then reads $|\psi_q\rangle := \exp(-i\delta\phi_q \hat{\mathcal{H}}) |\psi_0\rangle $. The denominator in Eq.~(6) (i.e. the norm of iterated wavefunction) is composed of overlaps $\mathcal{O}_q := \langle \psi_q | \psi_0 \rangle$, weighted by the $k$-dependent coefficients. The numerator is composed in the same way with modified overlaps $\mathcal{O}_q^{\mathcal{H}} := \langle \psi_q | \hat{\mathcal{H}}|\psi_0 \rangle$. The measurement procedure is detailed in methods (sec.~B), and for $\mathcal{O}_q^{\mathcal{H}}$ can follow the same schedule, where in the absence of $\hat{\mathcal{H}}$ operation the term-by-term estimation common to VQE approach can be used. We chosen the reference state to be $|\psi_{\mathrm{R}} \rangle = (\downarrow, \downarrow, \downarrow, \downarrow)^T$.
\begin{figure*}
\includegraphics[width=1.\textwidth]{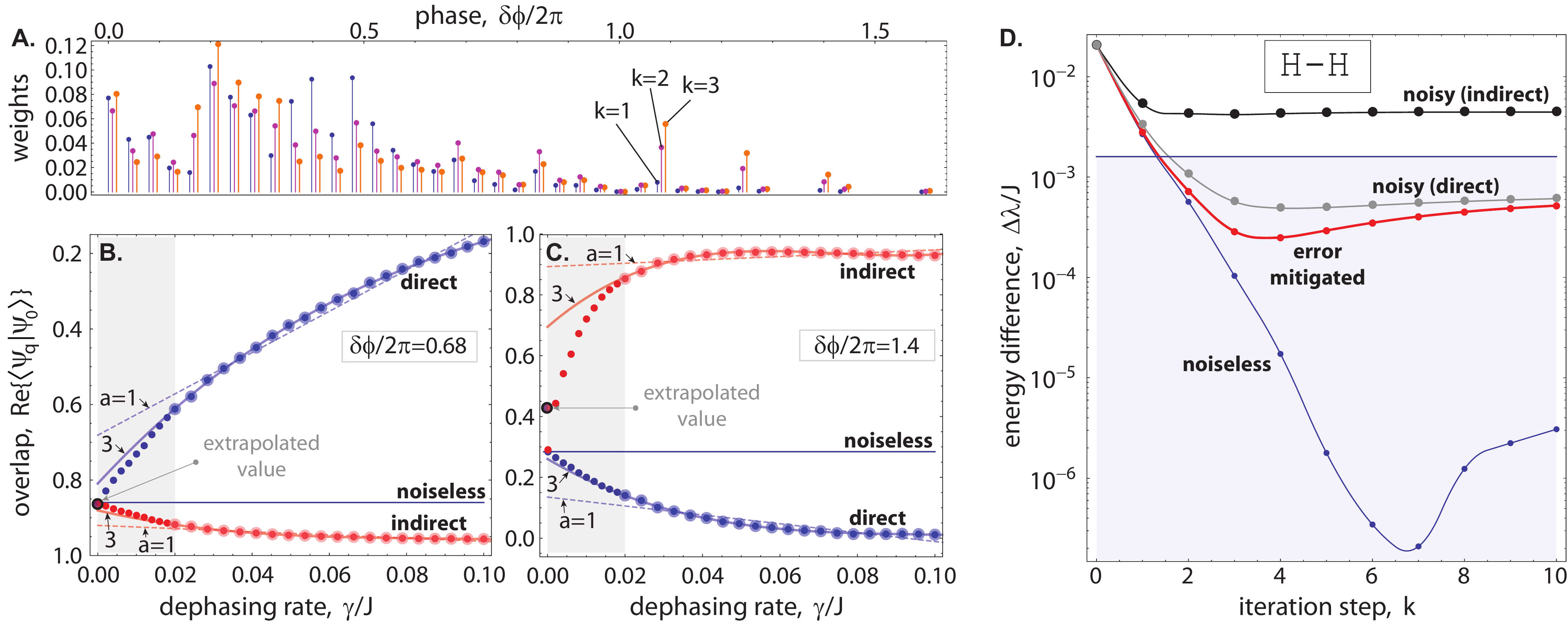}
\caption{Noise mitigation for molecular hydrogen (H--H). \textbf{A} Histogram of weights $p_{k,q}$ for different wavefunction overlaps generated by the evolution with phase differences $\delta\phi_q$. Approximation grid is fixed for all panels, and is set by $M_y = M_z = 5$, $\Delta_y J = \Delta_z = 0.5$, with $\phi_{\mathrm{max}}/2\pi \approx 1$. Blue, magenta, and red lines correspond to $k=1,2,3$ iteration steps, respectively. Each case is normalized by the total weight at given $k$. \textbf{B, C} Real parts of the wavefunction overlaps $\mathrm{Re}\{ \langle \psi_q| \psi_0 \rangle \} = \mathrm{Re}\{ \mathcal{O}_q \} $ as a function of the effective dephasing rate $\gamma$. The Monte-Carlo procedure was performed using 5000 trajectories for each point, and the phase difference was fixed to $\delta\phi_q/2\pi = 0.68$ (\textbf{B}) and $\delta\phi_q/2\pi = 1.4$ (\textbf{C}). Two types of overlap estimation are used, given by direct and indirect inference from Eqs.~(M4) and (M6) of Methods sec. B, respectively. The noiseless case of $\gamma = 0$ is shown by the horizontal blue line. The error mitigation is performed by using results of extrapolation of linear ($a=1$) and cubic ($a=3$) order to $\gamma = 0$ value. The result for the weighted procedure [see Eq.~\eqref{eq:O_avg}] is shown by the red circle and highlighted as \textit{extrapolated value}. Grey shaded region contains the results for small dephasing rate, and marks the inaccessible region excluded from the mitigation. \textbf{D} Combined results of quantum inverse iteration as a function of iteration step, shown for different data analysis strategies. Top black and grey curves correspond to noisy data with direct and indirect inference, respectively, taken at $\gamma_{\mathrm{min}}/J = 0.02$ ($\gamma_{\mathrm{min}}/\bar{\xi} \approx 0.16$). The red curve shows the error mitigated result. The lowest blue curve is for the noiseless operation for $\phi_{\mathrm{max}}/2\pi \approx 1$. In each case, the difference between true ground state and its estimate version is plotted (log scale), and blue shaded region denotes the chemical precision.}
\label{fig:noisy}
\end{figure*}

To account for the noisy operation we exploit the wavefunction Monte-Carlo (WFMC) approach well-known in quantum optics~\cite{Dalibard1992_SM}. Contrary to a mixed state description~\cite{LiBenjamin2017_SM,Temme2017_SM} this allows to operate in the original vector space. Importantly, it relates directly to an experimental workflow, and provides extra intuition for running the algorithm in noisy setting. In a somewhat modified form, it was already applied in Refs.~\cite{Intel2018_SM,Bassi2008_SM}. WFMC relies on calculating the evolution of the system subjected to noise in the form of quantum jump operator $\hat{C}_j$ for qubit $j$. Considering the uncorrelated dephasing processes for all qubits, we define jump operators as $\hat{C}_j = \sqrt{\gamma} Z_j$, where $\gamma$ is a dephasing rate. The expectation values are then measured over an ensemble of trajectories with different noise realization, and each probability overlap $|\mathcal{O}_{0,q}|^2,|\mathcal{O}_{+,q}|^2,|\mathcal{O}_{i,q}|^2$ is measured separately in the numerical procedure, mimicking the experiment.

We consider the concrete example of molecular hydrogen quantum inverse iteration with the Fourier approximation grid fixed to $M_y = M_z = 5$ and $\Delta_y J = \Delta_z = 0.5$, with $\phi_{\mathrm{max}}/2\pi \approx 1$. To perform energy estimation we consider explicitly all terms in Eq.~(6) of the main text and observe that in many cases the propagation phases coincide. In the following the unique phase difference values are chosen, and we are interested only in their absolute values, as $\pm \delta\phi_q$ evolution yields equal real parts for the overlap for systems which respect time-reversal symmetry. Finally, $\delta\phi = 0$ is accounted for trivially. This brings us to just 35 phase difference values to be used for quantum evolution. Next, the classical post-processing is performed, where measured noisy values for overlaps $\mathcal{O}_q$ and $\mathcal{O}_q^{\mathcal{H}}$ are multiplied by coefficients $p_{k,q}$ and summed together, taking different values of iteration step. An important information at this step is the prefactor in front of each overlap, as it gives a weight in the total estimate. Note that in the case of noisy operation this changes the success of the estimate, as it favors large weight for small $\delta\phi$ evolution and small weight for large $\delta\phi$'s.

In Fig.~\ref{fig:noisy}A we show the histogram of weights, taken as $w_{k,q} = (\sum_i |p_{k,q_i}|)/(\sum_{q,i} |p_{k,q_i}|)$ for different iteration step number $k$. Here, $i$ summation goes over all coefficients for the same $\delta\phi_q$, and each weight is normalized by the total sum of prefactors at given $k$. The histogram reveals that for small $k$ the main contribution comes from small-to-intermediate size phase differences $\delta\phi/2\pi <0.5$, while higher lying weights are nearly negligible. As iteration step $k$ grows, the estimates rely more on the overlaps for larger $\delta\phi$ (longer propagation times).

WFMC simulation was performed using $5000$ trajectories and considering different noise rates $\gamma$ ranging from $10^{-4}J$ to $0.1J$. Importantly, this is also compared to the average interaction constant for the Hamiltonian $\bar{\xi} = 0.1224 J$, taken as a norm of $\hat{\mathcal{H}}_{\mathrm{H}_2}$. The case of $\gamma \sim \bar{\xi}$ then corresponds to the highly noisy limit, and genuine operation typically requires $\gamma/\bar{\xi} \ll 1$. The noise rate can effectively be changed (increased) in the simulation. As $\bar{\xi}$ is generally tunable, it sets the physical timescale for the operation. Simultaneously rescaling the physical couplings $\{\xi_m\}$ by changing $J$ and physical evolution time, one can perform calculations for the same propagation phase $\delta\phi$ but with effectively different noise influence.

The examples of the measured overlap values are shown in Fig.~\ref{fig:noisy}B,C for phase differences $\delta\phi/2\pi = 0.68$ (B) and $\delta\phi/2\pi = 1.4$ (C). They were obtained using two approaches, which we refer as \emph{direct} and \emph{indirect} estimation. As explained in the Methods, sec. B, the direct estimation refers to measurement of $\mathrm{Re}\{\mathcal{O}_{0,t}\}$  using Eq.~(M4) and the indirect approach uses Eq.~(M5). The corresponding result for the direct measurement is shown in blue dots, and the indirect corresponds to red dots. In both cases, the measured estimate coincides with the noiseless value at $\gamma \rightarrow 0$, shown by the blue line (see Fig.~\eqref{fig:noisy}B,C).

Next, in the spirit of the error mitigation technique~\cite{LiBenjamin2017_SM,Temme2017_SM}, we choose an optimal point $\gamma_{\mathrm{min}}$ where the dephasing rate is minimized (overall coupling strength $J$ is maximized). This marks the best uncorrected results, and in the calculations we set $\gamma_{\mathrm{min}}/J = 0.02$, which is equal to $\gamma_{\mathrm{min}}/\bar{\xi} \approx 0.16$. Being only a one-sixth of the interaction strength, it certainly makes the measurement less feasible, and we choose it as an exaggerated case aiming to see if the quantum inverse iteration can deal with high-noise operation mode. Taking the measurements at effectively elevated noise rates with $\gamma > \gamma_{\mathrm{min}}$ (bold circles), the overlap is extrapolated to $\gamma/J = 0$ value. This is done using linear ($a=1$) and cubic ($a=3$) spline fitting, which corresponds to thin dashed and thick solid lines in Fig.~\ref{fig:noisy}B,C. The gray shaded area corresponds to the inaccessible range of noise rates.

We observe that for smaller propagation phase $\delta\phi/2\pi = 0.68$ (B) extrapolation of both direct and indirect measurement provides a decent estimates for the overlap $\mathcal{O}_{a=1,3}^{\mathrm{(dir)}}(0)$ and $\mathcal{O}_{a}^{\mathrm{(ind)}}(0)$ ($a=1,3$), with direct estimate giving more precise values. For $\delta\phi/2\pi = 1.4$ and long propagation time the number of errors grows, leading to the deviation of extrapolated values (Fig.~\ref{fig:noisy}C). The procedure is complicated by nonlinear overlap dependence. Again, direct estimation yields better estimate, although we note that this is not a universal trend. Having collected the information, we also compose the weighted averaged overlap value as
\begin{align}
\label{eq:O_avg}
\mathcal{O}_{a}^{\mathrm{(avg)}}(0) &= \frac{\mathsf{w}_{\mathrm{dir}}}{(\mathsf{w}_{\mathrm{dir}} + \mathsf{w}_{\mathrm{ind}})} \cdot \mathcal{O}_{a}^{\mathrm{(dir)}}(0) \\ \notag &+ \frac{\mathsf{w}_{\mathrm{ind}}}{(\mathsf{w}_{\mathrm{dir}} + \mathsf{w}_{\mathrm{ind}})}\cdot \mathcal{O}_{a}^{\mathrm{(ind)}}(0),
\end{align}
where $\mathsf{w}_{\mathrm{dir}} = [\mathcal{O}_{1}^{\mathrm{(ind)}}(0) - \mathcal{O}_{3}^{\mathrm{(ind)}}(0)]^2$ and $\mathsf{w}_{\mathrm{ind}} = [\mathcal{O}_{1}^{\mathrm{(dir)}}(0) - \mathcal{O}_{3}^{\mathrm{(dir)}}(0)]^2$. The reasoning behind the heuristic weighting procedure is as follows. First, the noise dependence for direct and indirect measurement of the overlap has opposite trends, with each deviating into higher or lower overlap values. This calls for a technique which can use the average of the two. Second, typically the cubic extrapolation with smaller curvature is more trustful; but when it largely deviates from linear approximation we enter the regime of high noise, and its likely that actual solution lies in between direct and indirect extrapolation result. The final result of the weighted mitigation is depicted in Figs.~\ref{fig:noisy}B,C by circles labelled as ``extrapolated value''.

Collecting the overlap data and coefficients, in Fig.~\ref{fig:noisy}D we plot the results of ground state estimation procedure for different post-processing techniques. The noiseless quantum inverse iteration is shown by the lowest (blue) curve, and serves as a reference. For noisy operation, the two upper curves correspond to overlap estimates \emph{without} error mitigation by taking their values at lowest achievable noise $\mathcal{O}^{\mathrm{(ind)}}(\gamma_{\mathrm{min}})$ and $\mathcal{O}^{\mathrm{(dir)}}(\gamma_{\mathrm{min}})$ for an indirect and a direct inference only. While the former stays outside of chemical precision for all iteration steps due to large overlap deviations (black curve), the direct inference technique allows to reach chemical precision even at high noise ($\gamma_{\mathrm{min}}/\bar{\xi} = 0.16$). Applying the error mitigation technique as stated in Eq.~\eqref{eq:O_avg}, the results can be improved for initial iteration steps, but approach unmitigated values for $k>10$. This can be explained by the change of the weight distribution for large $k$, where overlaps at large times are important (Fig.~\ref{fig:noisy}A), and weighted extrapolation does not provide good estimate in this case.

\end{document}